\documentclass{aa}  

\usepackage{array}
\usepackage{graphicx}
\usepackage{natbib}
\usepackage{lscape}
\usepackage{rotating}
\newcommand{\xmm}{{\sc XMM}\emph{-Newton}}

\newcommand{\tr}{Tr16-22}
\newcommand{\kms}{km\,s$^{-1}$}

\begin{document}

   \title{The puzzling properties of the magnetic O star Tr16-22\thanks{based on \xmm\ observations (ObsIDs 0691970101, 0742850301, 0742850401, 0762910401) and ESO data (Prog. 386.D-0624A, 086.D-0997B, 089.D-0975A, 091.D-0090B, 095.D-0082).}}


\author{Ya\"el Naz\'e\inst{1}\fnmsep\thanks{FNRS Research Associate}, Rodolfo Barb\'a\inst{2}, Stefano Bagnulo\inst{3}, Nidia Morrell\inst{4}, Roberto Gamen\inst{5}, V\'eronique Petit\inst{6}, Coralie Neiner\inst{7}}

\institute{Groupe d'Astrophysique des Hautes Energies - STAR, Institut d'Astrophysique et de G\'eophysique - B5c, Universit\'e de Li\`ege, 19c All\'ee du 6 Ao\^ut, B-4000 Sart Tilman, Belgium\\
\email{naze@astro.ulg.ac.be}
\and
Departamento de F\'{\i}sica y Astronom\'{\i}a, Universidad de La Serena, Av. Juan Cisternas 1200 Norte, La Serena, Chile
\and
Armagh Observatory and Planetarium, College Hill, Armagh, BT61 9DG, UK
\and
Las Campanas Observatory, Carnegie Observatories, Casilla 601, La Serena, Chile
\and
Instituto de Astrof\'isica de La Plata, CONICET--UNLP, and Facultad de Ciencias Astron\'omicas y Geof\'isicas, UNLP, Argentina
\and
Department of Physics and Space Sciences, Florida Institute of Technology, Melbourne, FL 32904, USA
\and
LESIA, Observatoire de Paris, PSL Research University, CNRS, Sorbonne Universit\'es, UPMC Univ. Paris 06, Univ. Paris Diderot, Sorbonne Paris Cit\'e, 5 place Jules Janssen, 92195 Meudon, France
             }

\authorrunning{Naz\'e et al.}
\titlerunning{More on Tr16-22}

 
  \abstract
   {The detection of bright, hard, and variable X-ray emission in \tr\ prompted  spectropolarimetric observations of this star, which in turn led to the discovery of a surface magnetic field.}
   {We want to further constrain the properties of this star, in particular to verify whether X-ray variations are correlated to changes in optical emission lines and magnetic field strength, as expected from the oblique rotator model that is widely accepted for magnetic O stars.}
   {We have obtained new low-resolution spectropolarimetric and long-term high-resolution spectroscopic monitoring of \tr, and we also analyse new, serendipitous X-ray data. }
   {The new X-ray observations are consistent with previous data, but their addition does not help to solve the ambiguity in the variation timescale because of numerous aliases. No obvious periodicity or any large variations are detected in the spectropolarimetric data of \tr\ obtained over three months. The derived field values appear to be in line with previous measurements, suggesting constancy of the field (though the possibility of small, short-term field variations cannot be excluded). Variations in the equivalent widths of H$\alpha$ are very small, and they do not appear to be related to the X-ray timescale; the overall lack of large variations in optical emission lines is consistent with the magnetic field constancy. In addition, variations of the radial velocities indicate that \tr\ is probably a SB1 binary with a very long period. }
   {Our new measurements of optical emission lines and magnetic field strength do not show an obvious correlation with X-ray variations. Our current data thus cannot be interpreted in terms of the common model, which assumes the electromagnetic emission associated with a wind confined by a dipolar field tilted with respect to the rotation axis. However, the sampling is imperfect and new data are needed to further constrain the actual periodicity of the various observed phenomena. If inconsistencies are confirmed, then we will need to consider alternative scenarios.}


   \maketitle
%

\section{Introduction}

Strong magnetic fields have been detected in a dozen O stars during the last decade \citep[and references therein]{pet13,fos15}. In such objects, the stellar wind flows are channelled towards the equator \citep{bab97}, creating a dense region of confined winds. Part of this material is shock-heated to higher temperatures, leading to the emission of X-rays, while the cooler plasma is easily detected in the visible (e.g. through  Balmer emission lines). When the rotation axis is not perfectly aligned with the magnetic axis,  a periodic modulation of the electromagnetic emission associated with these confined winds is observed.  Variations of the longitudinal field, of the broad-band photometry, of the visible emission lines, and of the X-ray flux thus occur simultaneously, as was found for example for $\theta^1$\,Ori\,C \citep{don02,gag05} or HD\,191612 \citep{don06,naz07,how07,naz10}.

\begin{table*}
 \centering
  \caption{List of the new X-ray observations. Phases were calculated considering $1/P$=0.01838\,d$^{-1}$ and $T_0$ as the date of the oldest \xmm\ observation (No. 4 in Table 1 of \citealt{naz14}, $JD$=2\,451\,751.707). }
  \label{list}
  \begin{tabular}{lcccc}
  \hline
ObsID & exp. & Start Date & Associated JD & $\phi$\\
& time & & & \\
  \hline
\hline
0691970101 & 87\,ks & 2012-Dec-20@19:21:54 & 2456282.307 & 0.27 \\
0742850301 & 13\,ks & 2014-Jun-06@19:13:05 & 2456815.301 & 0.07 \\
0742850401 & 33\,ks & 2014-Jul-28@15:32:43 & 2456867.148 & 0.02 \\
0762910401 & 11\,ks & 2015-Jul-16@01:18:44 & 2457219.555 & 0.50 \\
\hline
\end{tabular}
\end{table*}

Spectropolarimetric observations found \tr\ (O8.5V) to be strongly magnetic \citep{naz12carina,naz14}. In the X-ray range, \tr\ displays a bright, variable, and hard emission atypical of single, ``normal'' massive stars \citep{eva04,ant08,naz11,naz14}, but in line with theoretical expectations for confined winds \citep{nazsurvey}. The Fourier analysis of the X-ray data clearly indicated the presence of a periodicity; the  favoured value was $\sim$54d but many aliases were present, leaving some ambiguity on the actual period \citep{naz14}. Therefore, to better constrain its physical properties, we needed additional data of \tr.

In this paper, we analyse these new data. Section 2 presents the observations used in the study,  Sect. 3 provides the results, and Sect. 4 reports our conclusions.

\section{Observations}

\subsection{X-ray observations}
Being close to $\eta$\,Carinae, Tr16-22 is frequently observed at high energies. Since the observations presented in \citet{naz14}, three new \xmm\ exposures are available in the archives and an additional one was kindly provided by Dr Kenji Hamaguchi (Table \ref{list}). In these observations, the target only appears on the MOS2 camera: it falls outside the field of view of the other two cameras (because of dead CCDs in MOS1 or  the use of the small window mode in MOS1 or pn). These new data were reduced with SAS v14.0.0 using calibration files available in December 2015 and following the recommendations of the \xmm\ team\footnote{SAS threads, see \\ http://xmm.esac.esa.int/sas/current/documentation/threads/ }. Only the best quality data ($PATTERN$ of 0--12) were kept and background flares were discarded. A source detection was performed on each EPIC dataset using the task {\it edetect\_chain} on the 0.4--10.0\,keV energy band and for a likelihood of 10 to find the best-fit position of the target in each dataset. We note that the source is not bright enough to present pile-up. EPIC spectra were extracted using the task {\it especget} for circular regions of 35'' radius centred  on these best-fit positions and, for the backgrounds, at positions as close as possible to the target considering crowding and CCD edges. The spectra were finally grouped, using {\it specgroup}, to obtain an oversampling factor of five and to ensure that a minimum signal-to-noise ratio of three (i.e. a minimum of 10 counts) was reached in each spectral bin of the background-corrected spectra. 

The X-ray spectra were then fitted within Xspec v12.7.0 using the same models as in \citet{naz14}, i.e. absorbed optically thin thermal plasma models, i.e. $wabs \times phabs \times \sum apec$, with solar abundances \citep{and89}. The first absorption component is the interstellar column, fixed to $4.4\times10^{21}$\,cm$^{-2}$ (a value calculated using the colour  excess of the star and the conversion formula $5.8\times10^{21}\times E(B-V)$\,cm$^{-2}$ from \citealt{boh78}), while the second absorption represents additional (local) absorption. For the emission components, we considered either two or four temperatures, as in \citet{naz14}. Table \ref{xfits} yields the results of our X-ray fittings. Since the 2012 December exposure is long, we also extracted the X-ray lightcurve of \tr\ using the same regions as for spectrum extraction. Dedicated $\chi^2$ tests were then performed and did not reveal any significant variability of \tr\ during this observation.

\begin{table*}
  \caption{Results from the X-ray spectral fitting.}
  \label{xfits}
  \begin{tabular}{lccccccc}
  \hline
\multicolumn{8}{l}{\it A. Model $wabs \times phabs \times (apec+apec)$ }\\
ID & $norm_1$ & $norm_2$ & $\chi^2$(dof) & $F_{\rm X}^{obs}$         & $L_{\rm X}^{ISMcor}$ & $HR$ \\
   & \multicolumn{2}{c}{(cm$^{-5}$)}&    & erg\,cm$^{-2}$\,s$^{-1}$ & erg\,s$^{-1}$     &     \\
  \hline
  \hline
0691970101 & 1.45$\pm$0.18e-3 & 3.43$\pm$0.22e-4 & 0.92 (46) & 2.58$\pm$0.10e-13 & 1.78e32 & 0.52$\pm$0.04 \\
0742850301 & 1.07$\pm$0.39e-3 & 2.91$\pm$0.43e-4 & 0.40 (9)  & 2.12$\pm$0.20e-13 & 1.39e32 & 0.56$\pm$0.11 \\
0742850401 & 1.09$\pm$0.23e-3 & 2.77$\pm$0.26e-4 & 0.91 (26) & 2.05$\pm$0.13e-13 & 1.37e32 & 0.54$\pm$0.07 \\
0762910401 & 1.53$\pm$0.43e-3 & 3.06$\pm$0.50e-4 & 0.61 (9)  & 2.40$\pm$0.25e-13 & 1.78e32 & 0.47$\pm$0.10 \\
\hline
\multicolumn{8}{l}{\it B. Model $wabs \times phabs \times \sum_4 apec$ }\\
ID & $norm_1$ & $norm_3$ & $norm_4$ & $\chi^2$(dof) & $F_{\rm X}^{obs}$         & $L_{\rm X}^{ISMcor}$ & $HR$ \\
   & \multicolumn{3}{c}{(cm$^{-5}$)}&               & erg\,cm$^{-2}$\,s$^{-1}$ & erg\,s$^{-1}$     &    \\
  \hline
  \hline
0691970101 & 2.44$\pm$0.97e-3 & 3.63$\pm$0.47e-4 & 1.04$\pm$0.25e-4 & 0.87 (45) & 2.71$\pm$0.21e-13 & 1.71e32 & 0.58$\pm$0.07 \\
0742850301 & 1.51$\pm$2.11e-3 & 3.03$\pm$1.08e-4 & 9.13$\pm$5.32e-5 & 0.40 (8)  & 2.24$\pm$0.37e-13 & 1.31e32 & 0.65$\pm$0.22 \\
0742850401 & 2.90$\pm$1.25e-3 & 2.18$\pm$0.60e-4 & 1.26$\pm$0.27e-4 & 0.70 (25) & 2.42$\pm$0.20e-13 & 1.44e32 & 0.67$\pm$0.11 \\
0762910401 & 3.28$\pm$2.31e-3 & 3.06$\pm$1.15e-4 & 1.04$\pm$0.51e-4 & 0.82 (8)  & 2.57$\pm$0.36e-13 & 1.74e32 & 0.53$\pm$0.15 \\
\hline
\end{tabular}
\\
{\footnotesize For the model $wabs \times phabs \times (apec+apec)$, absorptions were fixed to $4.4\times10^{21}$\,cm$^{-2}$ and $6.3\times10^{21}$\,cm$^{-2}$, and temperatures to 0.28\,keV and 1.78\,keV (as in \citealt{naz14}, Table 2). For the model $wabs \times phabs \times \sum_4 apec$ (used for the general spectral fits in the X-ray survey of magnetic stars performed by \citealt{nazsurvey}), absorptions were fixed to $4.4\times10^{21}$\,cm$^{-2}$ and $7.0\times10^{21}$\,cm$^{-2}$, and temperatures to 0.2, 0.6, 1.0, and 4.0\,keV, while $norm_2$ was kept to zero (as in \citealt{naz14}, Table 3). The remaining free parameters, the strengths of the thermal components, are listed above with their $1\sigma$ errors, the goodness of fit, and the number of degrees of freedom. The last columns provide the observed fluxes and ISM-absorption corrected luminosities (both in the 0.5--10.\,keV range and for a distance of 2290\,pc), as well as the ratio ($HR$) between the hard (2.--10.0\,keV) and soft (0.5--2.0\,keV) ISM absorption-corrected fluxes.}
\end{table*}

\subsection{Spectropolarimetry}

Spectropolarimetric data of \tr\ had been obtained with the Very Large Telescope equipped with FORS2 in 2011 \citep{naz12carina} and 2013 \citep{naz14}. In view of the variations detected in X-rays, we requested a monitoring of the star (Prog. 095.D-0082, PI Naz\'e): seven additional observations were taken in service mode between April and June 2015 (Table \ref{Bval}). These new data were taken with the red CCD (a mosaic composed of two 2k$\times$4k MIT chips) without binning, with a slit of 1'' and the 1200B grating ($R\sim 1400$). The observing sequence consisted of 8 subexposures of 240s duration (except for one night, 2015 June 06, where it was pushed up to 500s) with retarder waveplate positions of $+45^{\circ}$, $-45^{\circ}$, $-45^{\circ}$, $+45^{\circ}$, $+45^{\circ}$, $-45^{\circ}$, $-45^{\circ}$, $+45^{\circ}$. We reduced these spectropolarimetric data with IRAF\footnote{http://iraf.noao.edu/ IRAF is distributed by the National Optical Astronomy Observatories, which are operated by the Association of Universities for Research in Astronomy, Inc., under cooperative agreement with the National Science Foundation.} as explained by \citet{naz12carina}: aperture extraction radius fixed to 20\,px, subtraction of nearby sky background, and wavelength calibration from 3675 to 5128\AA\ (with pixels of 0.25\AA) considering arc lamp data taken at only one retarder waveplate position (in our case $-45^\circ$). This allowed us to construct the normalized Stokes $V/I$ profile, as well as a diagnostic ``null'' profile \citep{don97,bag09}. We note that the signal-to-noise ratio of the derived $I$ spectra near 5000\AA\ was about 1100--1200. Finally, the associated longitudinal magnetic field was estimated by minimizing $\chi^2 = \sum_i \frac{(y_i - B_z\,x_i - a)^2}{\sigma^2_i}$ with $y_i$ either $V/I$ or the null profile at the wavelength $\lambda_i$ and $x_i = -g_\mathrm{eff}\ 4.67 \times 10^{-13} \ \lambda^2_i\ 1/I_i\ (\mathrm{d}I/\mathrm{d}\lambda)_i$ \citep{bag02}. This was done for $x_i$ in the interval between $-10^{-6}$ to $+10^{-6}$ (an interval where the vast majority of good points are available, avoiding potential problems in slope determination due to a few isolated data points at extreme $x_i$, although we note that enlarging the interval, e.g. $\pm1.6\times10^{-6}$, does not significantly change the reported values) after discarding edges and deviant points, after rectifying the Stokes profiles, and after selecting spectral windows centred on lines \citep[see][for further discussion]{naz12carina}. Table \ref{Bval} yields the resulting field values, along with the values for the old datasets (recalculated for the same spectral windows). 

\begin{table}
  \caption{Results from the spectropolarimetry (using rectification and within the same spectral windows). Phases were calculated as in Table \ref{list}.  }
  \label{Bval}
  \begin{tabular}{lcccc}
  \hline
Date & HJD & $\phi$ & $B_z$ (G) & $N_z$ (G) \\
& --2450000 & & & \\
\hline
\hline
2011-Mar-12 & 5632.617 & 0.33 & $-502\pm$77 & 38$\pm$79  \\
2011-Mar-13 & 5633.589 & 0.34 & $-662\pm$94 & $-89\pm$89 \\
2013-Apr-18 & 6400.581 & 0.45 & $-479\pm$79 & $-87\pm$73 \\
2013-Jul-29 & 6503.483 & 0.34 & $-543\pm$124& 43$\pm$121 \\
2015-Apr-03 & 7115.616 & 0.59 & $-529\pm$91 & 132$\pm$91  \\
2015-May-01 & 7143.590 & 0.10 & $-288\pm$88 & $-204\pm$83 \\
2015-May-11 & 7153.715 & 0.29 & $-189\pm$100& $-264\pm$96 \\
2015-May-19 & 7161.641 & 0.43 & $-530\pm$95 & $-61\pm$93  \\
2015-May-20 & 7162.626 & 0.45 & $-205\pm$92 & $-93\pm$85  \\
2015-Jun-06 & 7179.545 & 0.76 & $-446\pm$76 & 23$\pm$69   \\
2015-Jun-18 & 7191.525 & 0.98 & $-446\pm$115& 94$\pm$110  \\
\hline
\end{tabular}
\end{table}

\subsection{High-resolution spectroscopy}

Data used in this work correspond to spectra collected in three different observatories over a time span of eighteen years. Ten spectra covering 3600--6100\AA\ with a resolving power of 15000 were obtained with the REOSC \'echelle spectrograph\footnote{on long-term loan from the University of Li\`ege} attached to the 2.15m Jorge Sahade telescope at Complejo Astron\'omico El Leoncito (CASLEO, Argentina), during  1997, 1998, 2011, and 2015. Thirteen spectra covering 3450--9850\AA\ with a resolving power of 40000 were gathered with the \'echelle spectrograph attached to the 2.5m du Pont telescope at Las Campanas Observatory (LCO, Chile), in  2010, 2013, 2015, and 2016. For these CASLEO and LCO spectra, calibration lamp exposures were secured immediately before or after each star integration at the same sky position. Data were reduced and processed in a standard way using the usual IRAF routines. In addition, two spectra of \tr\ were obtained with the FEROS spectrograph attached to the 2.2m telescope at ESO La Silla Observatory (Chile) in  2011 (086.D-0997B) and 2015 (089.D-0975A). These spectra, covering 3570--9210\AA\ with a resolving power of 46000, were gathered following the standard ESO procedures, and they were reduced using the FEROS pipeline provided for MIDAS\footnote{http://www.eso.org/sci/software/esomidas/}. We note that spectra obtained after 2010 were taken in the framework of the program ``OWN Survey'', a spectroscopic monitoring of southern Galactic O- and WN-type stars \citep{bar10}, while the CASLEO spectra obtained in the nineties are part of the XMEGA project \citep{cor99,alb01}. 

\section{Results and discussion}

\begin{figure*}
\includegraphics[width=6cm]{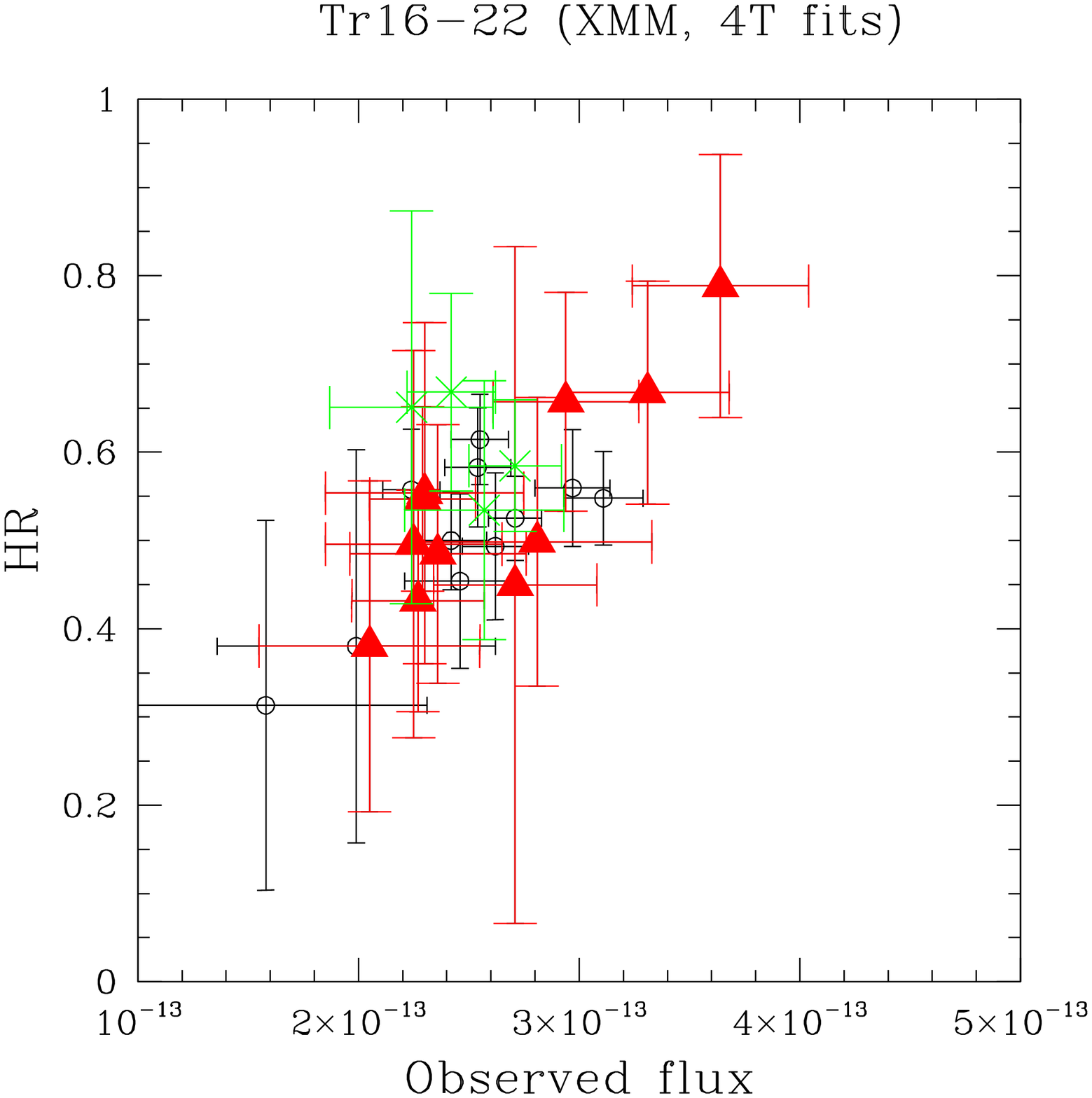}
\includegraphics[width=6cm]{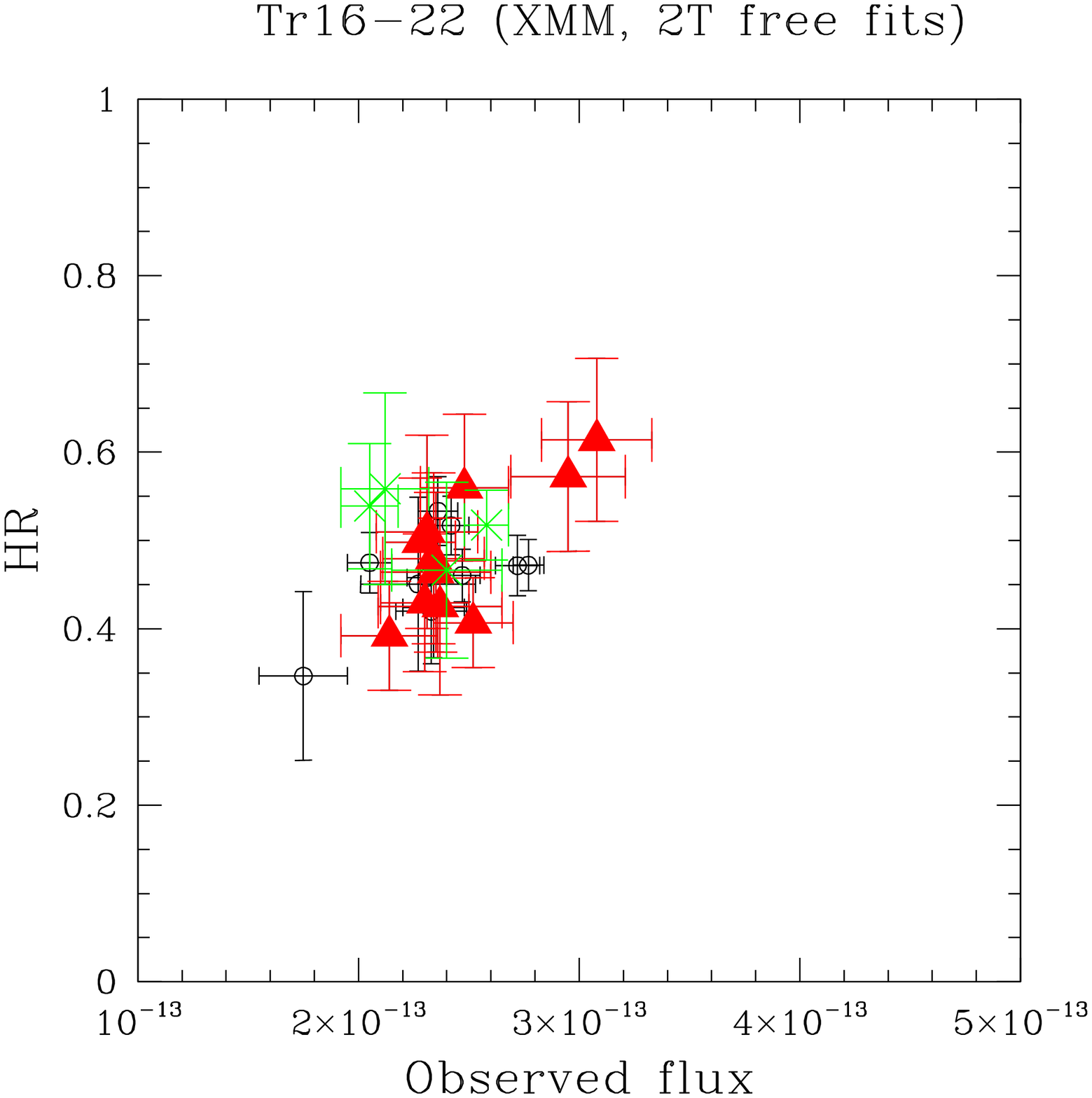}
\includegraphics[width=6cm]{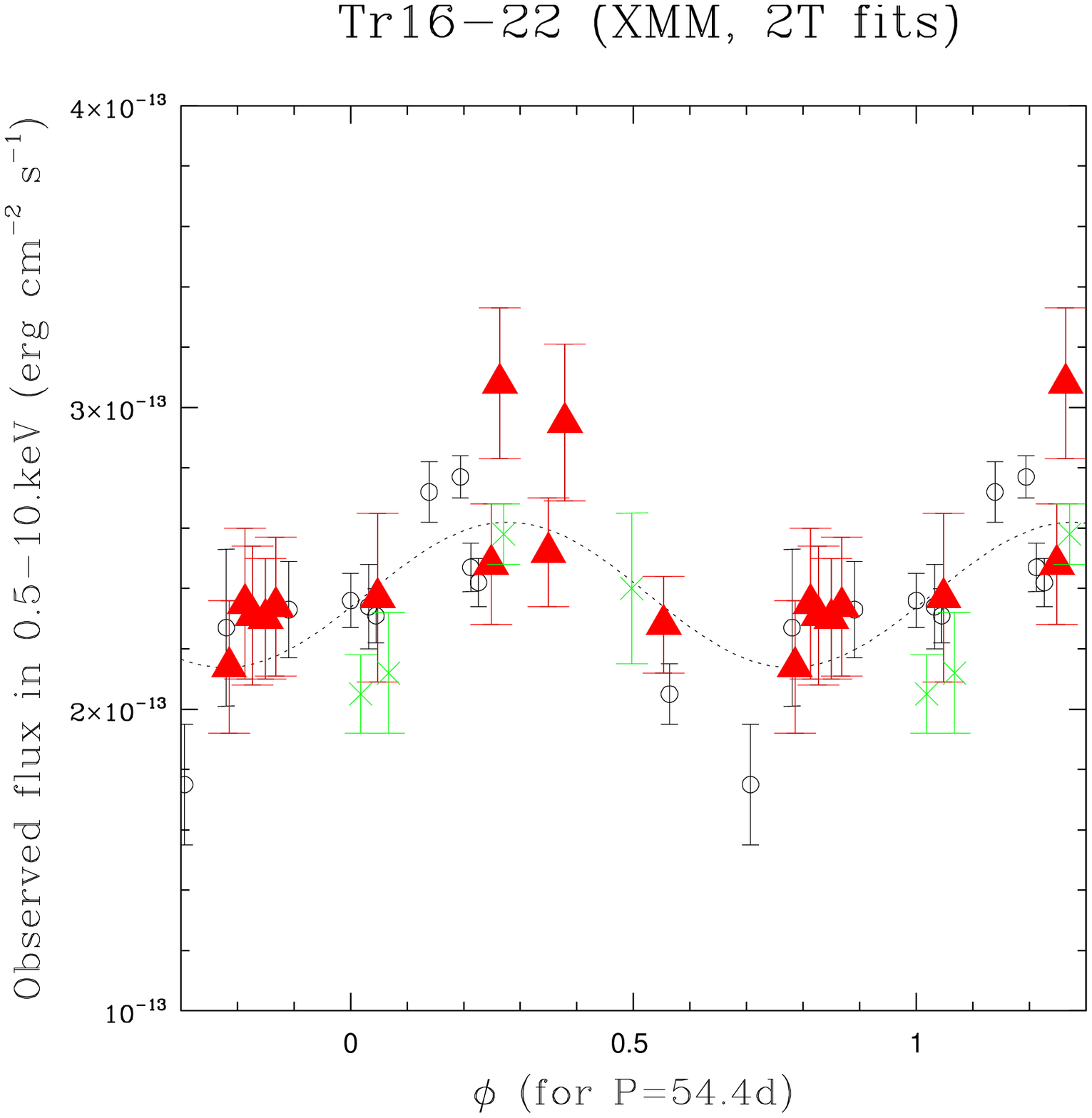}
\caption{{\it Left and Middle:} Evolution of the hardness ratios ($HR=H/S=F^{ISMcor}(2.-10.0\,keV)/F^{ISMcor}(0.5-2.\,keV)$) as a function of observed fluxes. Red triangles correspond to the 2003 \xmm\ data (the year with the largest number of observations), green crosses to the new data, and black open circles to the other \xmm\ observations. The values shown in the left (resp. middle) panel correspond to the results from 4T fits (resp. 2T fits). {\it Right:} Evolution of the observed fluxes from 2T fits as a function of phase (with $1/P$=0.01838\,d$^{-1}$ and $T_0$=2\,451\,751.707), along with the best-fit sinusoid (dotted line). }
\label{xvar}
\end{figure*}

In the X-ray range, the new data are in line with previous observations. First, the new data appear amongst the old ones in the flux-hardness relation (see Fig. \ref{xvar}). Second, the periodogram does not change much if the new data are included: $\sim$54\,d is still the favoured period and folding  with this timescale yields the same, coherent behaviour as shown before \citep[see right panel of Fig. \ref{xvar} and][]{naz14}. However, ambiguities on the period remain because aliases are still numerous (typically spaced by $4\times10^{-4}$\,d$^{-1}$). As mentioned in \citet{naz14}, the datasets recorded in 2003 provide a hint that the variation timescale may not be very long as the observed flux increased that year by about 30\% over two weeks at the end of July and beginning of August, then went back to its level in mid-August. Nevertheless, no full cycle was observed in a single observing run, and only observations with a higher cadence might be able to remove the ambiguity on the periodicity.

\begin{figure}
\includegraphics[width=8.5cm]{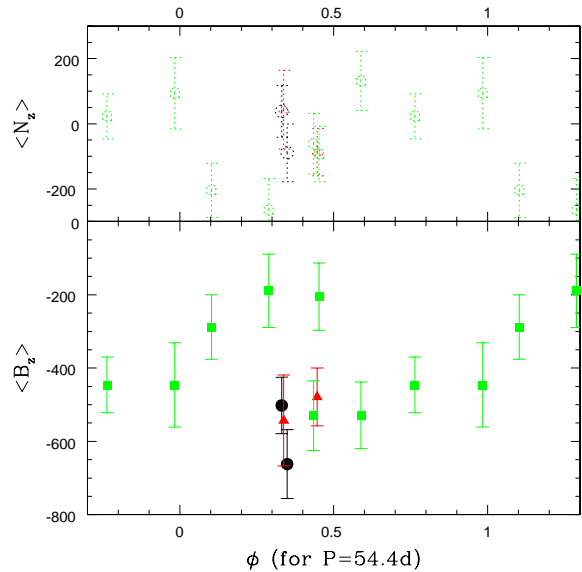}
\caption{Longitudinal field values (filled symbols) and the associated values for the null profiles (dotted empty symbols) as a function of phase (with $1/P$=0.01838\,d$^{-1}$ and $T_0$=2\,451\,751.707). The black circles correspond to 2011 data, the red triangles to 2013 data, and the green squares to 2015 data. }
\label{bval}
\end{figure}

Concerning magnetic field measurements, the 2015 FORS2 data formally provide only three secure detections (i.e. detections at $5-6\sigma$;  see \citealt{bag12}). Nevertheless, these new data  sample very different phases of the favoured X-ray timescale, allowing it to be tested (see phases in the second column of Table \ref{bval}). 
Unfortunately, the longitudinal field values do not show coherent variations when folded with this timescale (see Fig. \ref{bval}):  secure detections and non-detections are both notably found around phase 0.3. The  problem remains even when using only the 2015 data, where the precision of the ephemeris has less impact on the folding; for example, only the first of two consecutive days (May 19 and 20) provide a secure detection. Formally, a $\chi^2$ test finds a very small chance (0.6\%) for the $B_z$ values to be constant, but in fact the errors are underestimated -- only $5\sigma$ detections are secure with FORS2 spectropolarimetry when $3\sigma$ is usually sufficient \citep{bag12}. Indeed, the largest difference between the 2015 values actually only amounts  to $\sim$300G, i.e. around $3\sigma$. Therefore, we can conclude that, within the errors, the field has remained constant over the three months of observations, i.e. variations may be possible, but only with a very small amplitude ($<$300G). Furthermore, the 2015 longitudinal field strengths are similar to those based on 2011 or 2013 data, hinting at long-term constancy, though no clear conclusion can be drawn on possibly longer timescales as their sampling is far from perfect. For completeness, we have performed a period search\footnote{We applied several period search algorithms: (1) the Fourier algorithm adapted to sparse/uneven datasets \citep[a method rediscovered recently by \citealt{zec09} - these papers also note that the method of \citealt{sca82}, while popular, is not fully correct, statistically]{hmm,gos01}, (2) two different string length methods \citep{lafkin,renson}, (3) three binned analyses of variances (\citealt{whi44}; \citealt{jur71}, which is identical but with no bin overlap, to the ``pdm'' method of \citealt{ste78}; and \citealt{cuy87},  which is identical to the ``AOV'' method of \citealt{sch89}), and (4) conditional entropy \citep[see also \citealt{gra13}]{cin99,cin99b}. Each of these methods has its advantages and its drawbacks; the most reliable is the Fourier method, while the fastest  are usually  analyses of variances -- but the multiple identification of the same signal secure a detection.} on $B_z$ and $N_z$ values, but the derived periodograms are similar hence no significant periodicity can be identified for the stellar field measurements. These results appear somewhat at odds with the X-ray variations. In the context of an oblique rotator model, the flux doubling recorded at high energies suggested a high $i+\beta$ value for the dipolar field, with large changes in $B_z$ being expected. On the contrary, as large magnetic field variations over tens of days appear excluded by observations, this rather suggests either a pole-on configuration ($i\sim0$), an alignment of rotational and magnetic axes ($\beta\sim0$), or a very long rotation period.

\begin{figure*}
\begin{minipage}{7.5cm}
\includegraphics[width=7.5cm,bb=15 140 600 500,clip]{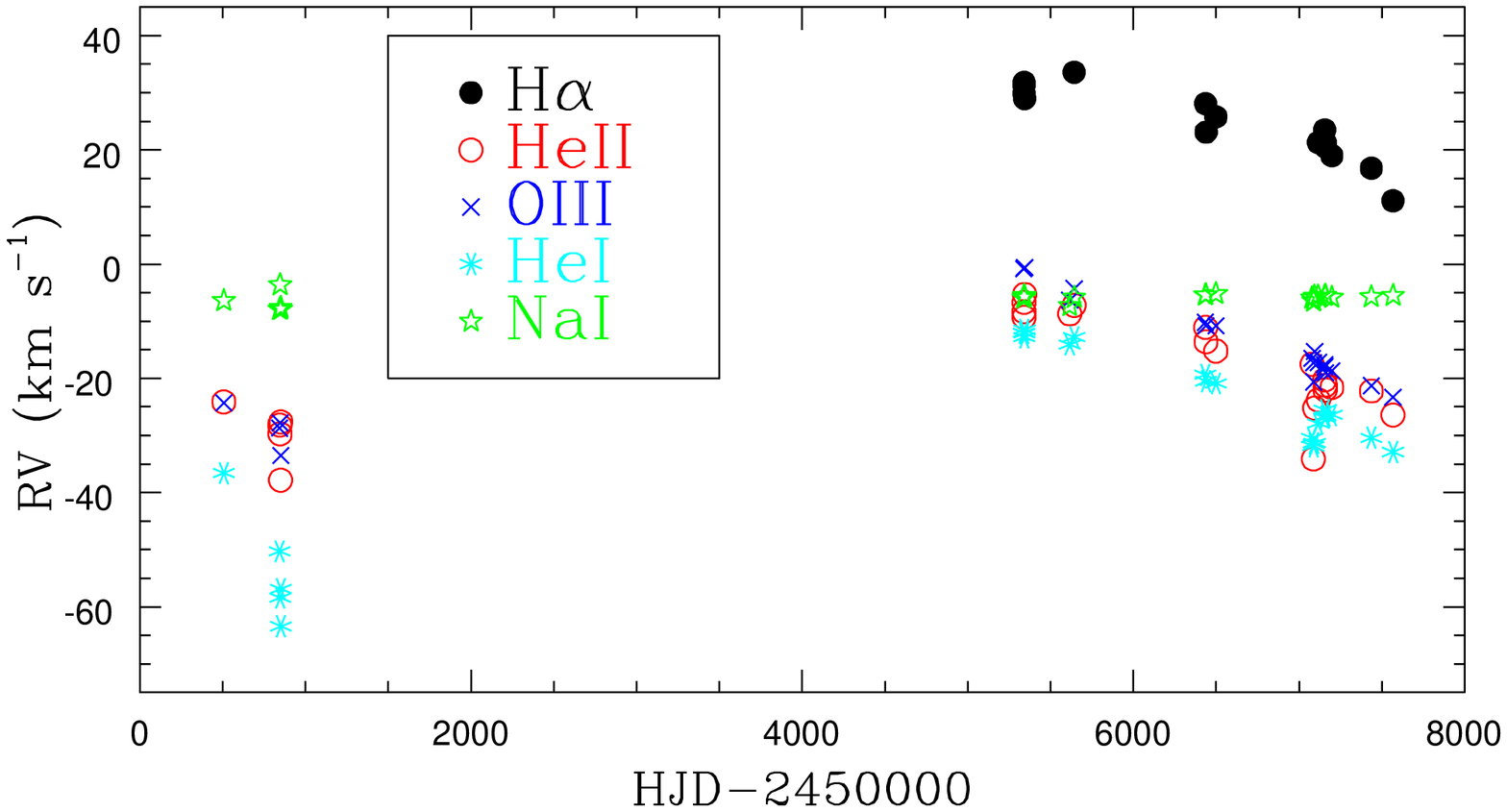}
\end{minipage}
\begin{minipage}{5.7cm}
\includegraphics[width=5.7cm]{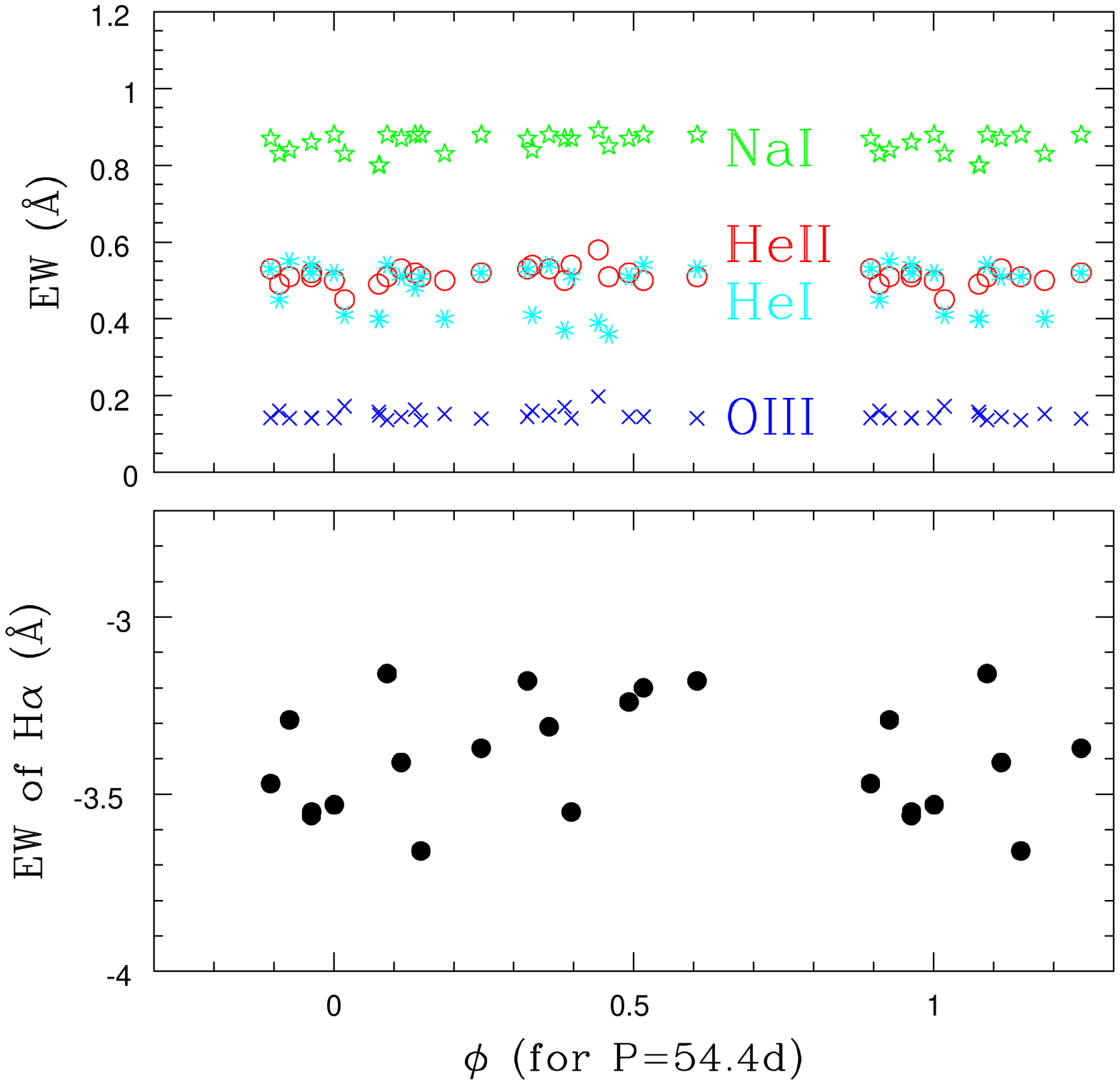}
\end{minipage}
\begin{minipage}{5.3cm}
\includegraphics[width=4.5cm,bb=100 115 475 575,clip, angle=-90]{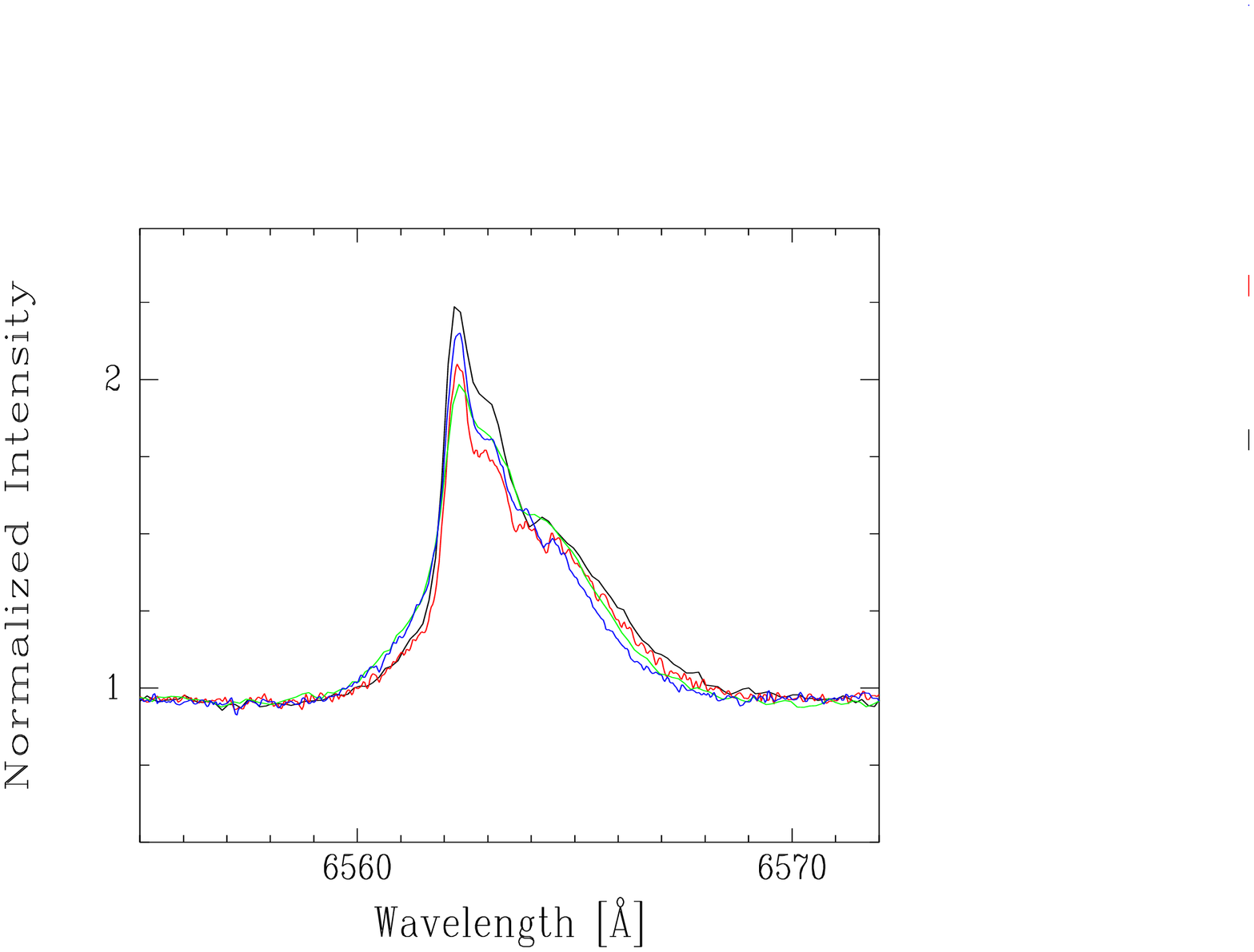}
\end{minipage}
\caption{{\it Left and Middle:} RVs (left panel) and EWs (middle panel) of selected lines (He\,{\sc ii}\,4686 shown as open red circles, O\,{\sc iii}\,5592 as blue crosses, He\,{\sc i}\,5876 as cyan asterisks, Na\,{\sc i}\,5890 as green stars, and H$\alpha$ as black dots) as a function of time or phase (with $1/P$=0.01838\,d$^{-1}$ and $T_0$=2\,451\,751.707), respectively. Typical errors are $<$5\,\kms\ for RVs and $<$10\% for EWs (see Table \ref{optique}). {\it Right:} H$\alpha$ profile observed in 2010 (black, $HJD$=2\,453\,40.538), 2011 (red, $HJD$=2\,455\,642.689),  2013 (green, $HJD$=2\,456\,498.481), and 2015 (blue, $HJD$=2\,457\,116.539). }
\label{ha}
\end{figure*}

For the optical spectra, the radial velocities and equivalent widths (EWs, estimated by integrating the line over a fixed interval) were measured for lines of different origins (interstellar, photospheric, circumstellar); the results are provided in Table \ref{optique}. For H$\alpha$, this was done after the narrow nebular component was subtracted, which could only be done  on the high-resolution spectra,  but this procedure is imperfect as demonstrated by e.g.  residual [O\,{\sc iii}] emissions. The radial velocities of all stellar lines yield coherent results (left panel of Fig. \ref{ha}). In particular, the systematic, coherent trend towards more negative velocities since 2010 (at a rate of $\sim$3.5\kms\ per year) should be noted. Formally, velocities are significantly variable: they display a larger dispersion than that of the interstellar Na\,{\sc i} line ($\sim$7\,\kms\ vs 0.2\,\kms\ in high-resolution data) and the maximum RV difference amount to about 20\,\kms\ corresponding to $\sim14\sigma$ since typical RV errors on high-resolution spectra are only 1\,\kms\ (i.e. RV variability criterion of \citealt{san13} is fulfilled). As no obvious line profile changes are detected, these RV variations indicate that \tr\ is a probable SB1 with a long period. To find the binary timescale, we applied the same set of period search algorithms as used before. Unfortunately, no  unambiguous periodicity could be pinpointed for the RVs measured on \tr, most probably because of the sparse sampling of the optical data. In view of the measured values, however, long periods ($>>$ tens of days, clearly incompatible with the 54\,d putative X-ray period) are favoured, and may be more precisely constrained by further monitoring the system. 

While detecting RV variations is an interesting result, our main objective was to study magnetically confined winds, and this is  performed through analysing line strengths of circumstellar emissions. It is now well known that Balmer lines, He\,{\sc ii}\,4686, and He\,{\sc i} lines are particularly sensitive indicators of confined winds \citep[see e.g. the case of HD\,191612 in ][]{naz07}. We found that, for any given line (interstellar, stellar, or circumstellar), the EWs always remain  similar, with a small dispersion ($<$0.01\AA\ on high-resolution data). Only the H$\alpha$ line displays some significant variations, and Fourier analyses tentatively yield a best period of about 8\,d for them. However, aliases are numerous and the current sampling of high-resolution spectra does not enable us to really test such a short period. Furthermore, folding magnetic field values or X-ray fluxes with this period results in a very dispersed graph, hence no clear timescale can be definitely identified. Finally, it should be noted  that the H$\alpha$ EW variations are of small amplitude (0.5\AA\ difference between the extreme EW values, Fig. \ref{ha}), an amplitude similar in magnitude to the 0.15\AA\ difference between extreme EW values recorded for the residual [O\,{\sc iii}]\,$\lambda$5007;  both EWs actually appear correlated, indicating some impact of the  nebular contamination (which varies because of the changing seeing) on the H$\alpha$ variations: the real H$\alpha$ variations are thus most probably of even smaller amplitude. These low-amplitude changes (or quasi constancy) seen in optical emission lines agree with the magnetic field results, but are at odds with the large variations detected in X-rays.

\begin{table*}
  \caption{Radial velocities (in \kms) and equivalent widths (in \AA) measured for some important lines. The third column indicates the instrument (F=FEROS@2.2mESO, L=\'echelle@LCO, C=REOSC@Casleo), while the fifth  provides the signal-to-noise ratios (S/N) of the spectra near 5000\AA. }
  \label{optique}
  \begin{tabular}{lcccccccccccccc}
  \hline
Date & HJD & Ins. & exp.  & S/N & \multicolumn{2}{c}{He {\sc ii} 4686} & \multicolumn{2}{c}{O {\sc iii} 5592} & \multicolumn{2}{c}{He {\sc i} 5876} & \multicolumn{2}{c}{Na {\sc i} 5890} & \multicolumn{2}{c}{H$\alpha$} \\
& $-$2.45e6& & time(s) & & RV & EW & RV & EW & RV & EW & RV & EW & RV & EW \\
\hline
\hline
1997-Feb-28 &  507.709 & C & 1800 &  50 &  $-$24.1  & 0.52  & $-$24.3 & 0.163 & $-$36.6 & 0.48 & $-$6.4 & 0.88&      & \\
1998-Jan-31 &  844.788 & C & 1800 &  60 &  $-$28.2  & 0.54  & $-$28.7 & 0.161 & $-$50.3 & 0.41 & $-$7.9 & 0.84&      & \\
1998-Feb-03 &  847.713 & C & 2400 &  70 &  $-$29.7  & 0.50  & $-$27.8 & 0.170 & $-$58.4 & 0.37 & $-$3.6 & 0.87&      & \\
1998-Feb-06 &  850.776 & C & 1800 &  50 &  $-$37.8  & 0.58  & $-$33.5 & 0.198 & $-$56.9 & 0.39 & $-$7.6 & 0.89&      & \\
1998-Feb-07 &  851.712 & C & 1800 &  40 &  $-$27.6  & 0.51  &         &       & $-$63.4 & 0.36 & $-$7.9 & 0.85&      & \\
2010-May-22 & 5338.547 & L & 1000 &  65 &  $-$8.3   & 0.51  & $-$0.8  & 0.141 & $-$12.0 & 0.55 & $-$5.9 & 0.84& 31.3 & $-$3.29 \\
2010-May-24 & 5340.538 & L & 1200 &  90 &  $-$6.7   & 0.51  & $-$0.7  & 0.142 & $-$12.8 & 0.54 & $-$5.6 & 0.86& 29.9 & $-$3.56 \\
2010-May-24 & 5340.552 & L & 1200 &  95 &  $-$9.3   & 0.52  & $-$0.7  & 0.141 & $-$12.1 & 0.52 & $-$5.5 & 0.86& 31.8 & $-$3.55 \\
2010-May-26 & 5342.608 & L & 1200 &  90 &  $-$5.3   & 0.50  & $-$0.6  & 0.142 & $-$11.5 & 0.52 & $-$5.7 & 0.88& 29.0 & $-$3.53 \\
2011-Feb-23 & 5615.579 & C & 1800 &  55 &  $-$8.7   & 0.45  & $-$6.3  & 0.172 & $-$14.1 & 0.41 & $-$7.3 & 0.83&      & \\
2011-Mar-22 & 5642.689 & F & 1800 &  90 &  $-$7.2   & 0.50  & $-$4.3  & 0.145 & $-$12.8 & 0.54 & $-$5.8 & 0.88& 33.6 & $-$3.2  \\
2013-May-23 & 6435.545 & L & 1500 & 100 &  $-$11.1  & 0.51  & $-$10.2 & 0.136 & $-$19.4 & 0.54 & $-$5.4 & 0.88& 28.1 & $-$3.16 \\
2013-May-26 & 6438.603 & L & 1500 & 100 &  $-$13.6  & 0.51  & $-$10.8 & 0.136 & $-$20.5 & 0.51 & $-$5.4 & 0.88& 23.1 & $-$3.66 \\
2013-Jul-24 & 6498.481 & L & 1200 &  90 &  $-$15.2  & 0.52  & $-$10.8 & 0.140 & $-$20.9 & 0.52 & $-$5.2 & 0.88& 25.8 & $-$3.37 \\
2015-Feb-25 & 7078.654 & C & 1800 &  40 &  $-$17.5  & 0.49  & $-$16.5 & 0.161 & $-$30.4 & 0.45 & $-$5.9 & 0.83&      & \\
2015-Mar-06 & 7087.669 & C & 1800 &  60 &  $-$34.1  & 0.49  & $-$17.3 & 0.158 & $-$31.9 & 0.40 & $-$6.4 & 0.80&      & \\
2015-Mar-06 & 7087.730 & C & 1800 &  50 &           &       & $-$20.6 & 0.149 & $-$31.4 & 0.40 & $-$6.5 & 0.80&      & \\
2015-Mar-12 & 7093.646 & C & 1800 &  60 &  $-$25.2  & 0.50  & $-$15.3 & 0.152 & $-$31.3 & 0.40 & $-$5.6 & 0.83&      & \\
2015-Apr-04 & 7116.539 & F & 2000 & 110 &  $-$23.8  & 0.51  & $-$17.2 & 0.141 & $-$27.9 & 0.53 & $-$5.6 & 0.88& 21.3 & $-$3.18 \\
2015-May-13 & 7155.553 & L & 1200 &  70 &  $-$20.3  & 0.53  & $-$17.6 & 0.145 & $-$25.5 & 0.53 & $-$5.5 & 0.87& 23.5 & $-$3.18 \\
2015-May-15 & 7157.514 & L & 1800 &  80 &  $-$22.0  & 0.53  & $-$18.0 & 0.148 & $-$26.9 & 0.54 & $-$5.5 & 0.88& 21.3 & $-$3.31 \\
2015-May-17 & 7159.537 & L & 1500 &  70 &  $-$21.2  & 0.54  & $-$19.0 & 0.141 & $-$26.6 & 0.51 & $-$5.5 & 0.87& 20.5 & $-$3.55 \\
2015-Jun-25 & 7198.523 & L & 1800 &  70 &  $-$21.6  & 0.53  & $-$18.7 & 0.144 & $-$26.4 & 0.51 & $-$5.8 & 0.87& 19.0 & $-$3.41 \\
2016-Feb-18 & 7436.794 & L & 1200 &  55 &  $-$22.2  & 0.52  & $-$21.3 & 0.144 & $-$30.4 & 0.51 & $-$5.7 & 0.87& 16.8 & $-$3.24 \\
2016-Jun-28 & 7567.519 & L & 1800 &  65 &  $-$26.4  & 0.53  & $-$23.3 & 0.142 & $-$32.9 & 0.53 & $-$5.5 & 0.87& 11.1 & $-$3.47 \\
\hline
\end{tabular}
{\footnotesize The systematic differences in CASLEO EW measurements of the He {\sc i} 5876 line are probably produced by the scattered light in the background (not perfectly subtracted).  For this detector,  we also note  that this line and the neighbouring interstellar lines appear at the edge of the detector, hence they are less reliable. There are some values missing for CASLEO data: the H$\alpha$ line is outside the range of the spectrograph and, in a few other cases, lines were too noisy to be reliable hence their measurements are not quoted here. Typical errors on RVs are about 1\,\kms\ for LCO and FEROS data and 3--5\,\kms\ for CASLEO data, while those on EWs are below 5\% in the LCO and FEROS spectra and between 5 and 10\% for the CASLEO spectra. }
\end{table*}

\section{Conclusion}
Spectropolarimetric, X-ray, and high-resolution spectroscopic data of \tr\ have been obtained. They show a good agreement with previous results: the X-ray properties are in line with those already reported and the magnetic field values are similar (within errors) to previous measurements. However, there are also surprises. 

The optical spectroscopy indicates the presence of RV changes in \tr, suggesting that it is a long-term SB1, but the sparse sampling of the optical data prohibits us from deriving a full orbital solution. The probable binary nature of \tr\ cannot explain its exceptional X-ray emission, however. First, colliding-wind emission in the system could be seen as one possible source of X-ray variability, but (1) the long timescale of the RV variations may be difficult to reconcile with the shorter timescale detected in X-ray data and (2) such late-type O stars do not exhibit bright colliding-wind X-ray emission. Second,  a low-mass (PMS) companion could be envisaged as the source of the X-ray variations because such objects display X-ray flares, but (1) the observed variations are large (about a factor of 2 corresponding to an increase in flux of $>10^{32}$\,erg\,s$^{-1}$),   an extreme value for typical PMS  flares, (2) PMS flares occur on relatively short timescales while the 2003 data indicate that a month was needed to return to ``normal'' flux levels, and (3) the combination of a long period and a large RV amplitude (at least 20\,\kms) of the O-type star rule out a companion with very low mass. Finally, it should  also be noted that the average luminosity of \tr\ ($\log[L_{\rm X}]\sim32.3$) is similar to expectations from confined wind models (in particular, see the right panel of Fig. 6 in \citealt{nazsurvey} where \tr\ is \#8).

On the other hand, neither the magnetic field values nor the optical spectroscopy display an obvious modulation with the favoured (54\,d) timescale derived from X-rays. Moreover,  the measured magnetic field is compatible with a constant value, or at most with a low-amplitude modulation, while the optical line strengths also remain remarkably constant in our data. This near-constancy is reminiscent of HD\,148937 \citep{naz08,wad12,nazsurvey}. It would imply a system with a long period, always seen close to pole-on, or with magnetic and rotational axes aligned, but in these latter cases the X-ray emission of the confined winds would also remain stable, as significant occultation of the X-ray emitting regions by the stellar body would not occur, and this clearly contradicts the X-ray observations of \tr. However, many aliases were present in the X-ray flux periodogram, rendering  the choice of the best period difficult and the new data did not fully resolve this ambiguity. 

The new data have thus brought additional questions. The variation timescales of the X-ray emission, optical emission line strength, and longitudinal magnetic field remain poorly constrained and -- worse -- their correlated behaviour has not yet been established, while it is clearly seen in all other magnetic O stars. The near-constancy of magnetic field values and optical emission strengths may even be difficult, if not impossible, to reconcile with the much larger variations detected at high energies. Only additional optical and X-ray observations, carefully scheduled and with high signal-to-noise ratios, would be able to clarify the situation and firmly confirm whether there is a mismatch for \tr\  with the usual oblique rotator scenario for confined winds in massive stars. 

\begin{acknowledgements}
The authors acknowledge Dr Nolan Walborn for fruitful discussions, Dr Kenji Hamaguchi for sharing his private X-ray data, as well as Francesco Di Mille and Claudio Germana who were observing on the du Pont telescope and kindly obtained the Feb. 2016 observation under NM's request. YN acknowledges support from  the Fonds National de la Recherche Scientifique (Belgium), the Communaut\'e Fran\c caise de Belgique, the PRODEX XMM contract (Belspo), and an ARC grant for concerted research actions financed by the French community of Belgium (Wallonia-Brussels federation). RHB is grateful for financial support from FONDECYT Regular Project No. 1140076.  RG was supported by grant PIP 112-201201-00298 (CONICET). RG and NM were visiting Astronomer, Complejo Astron\'omico El Leoncito operated under agreement between the Consejo Nacional de Investigaciones Cient\' ificas y T\'ecnicas de la Rep\'ublica Argentina and the National Universities of La Plata, C\'ordoba, and San Juan.
\end{acknowledgements}

\end{document}